
\documentstyle{amsppt}
\magnification=1200
\baselineskip=18pt
\nologo
\TagsOnRight
\document
\define \ts{\thinspace}
\define \oa{{\frak o}}
\define \g{{\frak g}}
\def\n{{\frak n}}
\def\h{{\frak h}}
\def\gl{{\frak {gl}}}
\def\aa{{\frak a}}
\define \Sym{{\frak S}}
\define \spa{{\frak {sp}}}
\define \U{{\operatorname {U}}}
\define \Z{{\operatorname {Z}}}
\define \Y{{\operatorname {Y}}}
\define \C{{\Bbb C}}
\define \E{{\Cal E}}
\define \si{{\Cal C}\ts}
\define \tsi{\tilde{\Cal C}\ts}
\define \sgn{\text{{\rm sgn}}}
\define \inv{\text{{\rm inv}}}
\define \ot{\otimes}
\define \qdet{\text{{\rm qdet}\ts}}
\define \sdet{\text{{\rm sdet}\ts}}
\define \End{\text{{\rm End}\ts}}

\define \ra{\rightarrow}

\define \Proof{\noindent {\bf Proof. }}

\heading{\bf YANGIANS AND CLASSICAL} \endheading
\heading{\bf LIE ALGEBRAS} \endheading
\bigskip
\heading{\bf Part II.\ \ \  Sklyanin determinant,} \endheading
\heading{\bf Laplace
operators and characteristic identities} \endheading
\bigskip
\heading{Alexander Molev}
\endheading
\bigskip
\heading{\rm Mathematics Research Report No. MRR 024-94}
\endheading
\bigskip
\noindent
Centre for Mathematics and its Applications \newline
School of Mathematical Sciences \newline
Australian National University \newline
Canberra, ACT 0200, Australia \newline
(e-mail:\ molev\@pell.anu.edu.au)

\bigskip
\noindent
{\bf May 1994}

\bigskip
\noindent
{\bf Mathematics Subject Classifications (1991).} 17B35, 17B37, 81R50

\bigskip
\noindent
{\bf Abstract}\newline
We study the structure of quantized enveloping algebras
called twisted Yangians, which
are naturally associated with the $B$, $C$, and $D$ series of
the classical Lie algebras. We obtain an explicit formula for
the formal series (the Sklyanin determinant) whose coefficients
are free generators of the center of the twisted Yangian.
As a corollary we obtain a new system of
algebraically independent generators of the center of the
universal enveloping algebra for the orthogonal and symplectic
Lie algebras and find the characteristic polynomial
for the matrix formed by the generators of these Lie algebras.

\newpage
\heading
{\bf 0. Introduction}
\endheading
\

Part I of the preprint (see Molev--Nazarov--Olshanski\u\i\ [MNO])
contains a detailed exposition
of some results describing the algebraic structure
of `quantum' algebras called Yangians and twisted Yangians.
The Yangian $\Y(\aa)$ is a canonical
deformation of the universal enveloping algebra for the polynomial
current Lie algebra $\aa[x]$ corresponding to a simple
complex Lie algebra $\aa$; see Drinfeld [D].
The algebra $\Y(N)=\Y(\gl(N))$, which may be called the Yangian for
the reductive Lie algebra $\gl(N)$, has been studied before in the works of
mathematical physicists from St.-Petersburg; see for instance
Takhtajan--Faddeev [TF].

The Yangian $\Y(N)$ is defined as a complex associative algebra
with quadratic defining relations involving the Yang $R$-matrix,
a rational solution of the quantum Yang--Baxter equation.
The algebraic
structure of $\Y(N)$ is studied in [MNO] by using the so-called $R$-matrix
formalism based on the properties of solutions of the Yang--Baxter
equation; see Takhtajan--Faddeev [TF], Kulish--Sklyanin [KS],
Reshetikhin--Takhtajan--Faddeev [RTF].
In particular, this approach enables one to construct a
system of free generators of the center of the algebra $\Y(N)$.
These generators appear to be the coefficients of a formal series
called the quantum determinant; see Kulish--Sklyanin [KS].

The Yangian $\Y(N)$ contains the universal
enveloping algebra $\U(\gl(N))$ as a subalgebra and one has a
homomorphism
$$
\Y(N)\ra\U(\gl(N)) \tag 1
$$
identical on $\U(\gl(N))$.
Using this homomorphism and the properties of the quantum determinant
one can construct the system of free
generators of the center of $\U(\gl(N))$
(see Nazarov--Tarasov [NT]) previously found
in Howe [H], Howe--Umeda [HU] and is closely related to
the Capelli identities [Ca1], [Ca2], [W].
Moreover, it was proved by Nazarov--Tarasov [NT]
that these generators actually
coincide with the coefficients of the polynomial identity
satisfied by the matrix $E$ formed by the basis elements of the Lie algebra
$\gl(N)$; cf. Bracken--Green [BG], Green [Gr],
O'Brien--Cant--Carey [BCC], Gould [G].

The properties of the quantum determinant
can be also applied to constructing the Gelfand--Tsetlin bases in
representations of the Yangian $\Y(N)$ and of the Lie algebra $\gl(N)$;
see Cherednik [C], Nazarov--Tarasov [NT], Molev [M].

This approach cannot be applied directly for the case of
the Lie algebra
$\aa$ not being of the $A$ series, because there is no
homomorphism of the Yangian $\Y(\aa)$ into the universal enveloping algebra
$\U(\aa)$ analogous to (1); see [D].
For the classical Lie algebras $\aa=\oa(N)$ and $\spa(N)$,
G. I. Olshanski\u\i\ [O2] introduced new algebras $\Y^+(N)$ and $\Y^-(N)$
respectively, called the twisted Yangians.
Each of these algebras is a deformation of the universal enveloping
algebra for a certain twisted polynomial current Lie algebra and can be
realized as a subalgebra in the Yangian $\Y(N)$.
Moreover, it was shown in [O1], [O2] that both the Yangian and
the twisted Yangian arise in conventional representation theory as
`true' analogues of the universal enveloping algebras of the
infinite dimensional classical Lie algebras $\frak u(\infty)$,
$\oa(\infty)$, $\spa(\infty)$.

As in the case of $\Y(N)$, the algebra $\Y^{\pm}(N)$
contains the universal enveloping algebra $\U(\aa)$ as a subalgebra and
one has the natural homomorphism
$$
\Y^{\pm}(N)\ra \U(\aa) \tag 2
$$
identical on $\U(\aa)$. The $R$-matrix formalism can be applied to the
investigation of the structure of the twisted Yangian $\Y^{\pm}(N)$ as
well;
see [O2], [MNO]. In particular, an analogue of the quantum determinant can
be
constructed. In [MNO] it was called the Sklyanin
determinant. It is a formal series whose coefficients are generators of the
center of the algebra $\Y^{\pm}(N)$. However, the formula for the
Sklyanin determinant, which can be derived directly from the
definition, is rather complicated because of the large number of
summands (it is $2^{N(N-1)/2}$ times longer than
the formula for the quantum
determinant), and so this restricts its possible applications.

In this paper we find a `short' formula for the
Sklyanin determinant analogous to that of the quantum determinant.
It involves a projection of the symmetric group $\Sym_N$ to
the subgroup $\Sym_{N-1}$, and we discuss some of its
combinatorial properties.
Then, using this formula and
the homomorphism (2), we construct a new
system of free generators of the center of the universal enveloping algebra
for
the orthogonal and symplectic Lie algebras.

A description of the center of the universal enveloping algebra of
a semi-simple Lie algebra $\g$ by means of invariant elements in
the symmetric algebra of a Cartan subalgebra in $\g$ is provided by
the Harish-Chandra isomorphism.
The important problem is to construct central elements explicitly in
terms of generators of the Lie algebra $\g$ and to find their
images with respect to the Harish-Chandra isomorphism.
A solution of this problem for the classical Lie algebras
was obtained by Perelomov--Popov [PP]. They constructed central
elements and found the corresponding invariant polynomials which
turned out to be
expressed in terms of sums of powers of variables.

The construction of central elements for the case of
orthogonal and symplectic Lie algebras
presented in this paper is analogous to the one mentioned above
for the Lie algebra $\gl(N)$ (see Howe--Umeda [HU, Appendix 1]). Here
the corresponding invariant polynomials turn out to be elementary symmetric
functions. Note that the family of central elements found by
Howe--Umeda [HU, Appendix 2] for the
orthogonal Lie algebra exhibits some similarity to ours. It would be very
interesting to find out what the connection is between these families.

In the first of the two constructions
it is not very difficult to prove that the elements
considered by Perelomov--Popov [PP] are central whereas
the calculation of their images under the Harish-Chandra isomorphism
is rather complicated. In the second construction, however, the most
difficult part is to verify that the elements belong to the center.
In our approach the use of the Yangian and the twisted Yangian
is crucial.

Furthermore, using the approach of Nazarov--Tarasov [NT]
we apply the formula for the Sklyanin determinant
to find the characteristic polynomial for the matrix $F$
formed by the generators of the orthogonal or
symplectic Lie algebras. Various properties of
the characteristic polynomials and the corresponding
identities for semi-simple Lie algebras were studied
by O'Brien--Cant--Carey [BCC] and Gould [G]. In particular,
the existence of the polynomial identities satisfied by the
matrices $E$ and $F$ was established. However,
their explicit form was not found (some low dimensional cases
were considered in Gould [G]).

Finally, we prove that the image
of the corresponding characteristic identities in a highest weight
representation coincides with the Bracken--Green identities [BG], [Gr].

This paper was written in a close cooperation with
Grigori Olshanski\u\i\ and Maxim Nazarov. I express my deep gratitude
to both of them.

\newpage
\heading
{\bf 1. The Yangian $\Y(N)$ and the quantum determinant}
\endheading
\

Here we reproduce some results exposed in [MNO] about the structure
of the algebra $\Y(N)$ (Subsections 1.1--1.4)
and then following the paper Nazarov--Tarasov [NT] we apply them to
the construction of a system of free generators of the center
of the universal enveloping algebra $\U(\gl(N))$ (see also
Howe--Umeda [HU]) and to finding
the characteristic polynomial for the matrix formed by the generators of
the Lie algebra $\gl(N)$ (Subsections 1.5--1.7).

\bigskip
\noindent
{\bf 1.1.} The {\it Yangian} $\Y(N)=\Y(\gl(N))$ is the
complex associative algebra with the
generators $t_{ij}^{(1)},t_{ij}^{(2)},\dots$ where $1\leq i,j\leq N$,
and the defining relations
$$
[t_{ij}^{(r+1)},t_{kl}^{(s)}]-
[t_{ij}^{(r)},t_{kl}^{(s+1)}]=
t_{kj}^{(r)}t_{il}^{(s)}-t_{kj}^{(s)}t_{il}^{(r)},
\tag1
$$
where $r,s=0,1,2,\ldots\;$ and $t_{ij}^{(0)}:=\delta_{ij}$.

For any $i,j=1,\ldots, N$ define the formal power series
$$
t_{ij}(u) = \delta_{ij} + t^{(1)}_{ij} u^{-1} + t^{(2)}_{ij}u^{-2} +
\cdots \in \Y(N)[[u^{-1}]] \tag2
$$
and combine these series into a single $T$-{\it matrix}:
$$
T(u):=\sum^N_{i,j=1} t_{ij}(u)\ot E_{ij}
\in \Y(N)[[u^{-1}]]\ot \End\E, \tag3
$$
where $\E:=\C^N$ and $E_{ij}$ are the standard matrix units.
We shall need to consider the multiple tensor products
$\E\ot\dots\ot\E$
and operators
therein.
For an operator $X\in \End\E$ and a number $m=1,2,\ldots$ we set
$$
X_k:=1^{\otimes (k-1)}\otimes X\otimes 1^{\otimes (m-k)}
\in\End\E^{\otimes m}, \quad 1\leq k\leq m. \tag4
$$
If $X\in\End\E^{\otimes 2}$ then for any $k, l$\  such that
$1\leq k, l\leq m$\ and $k\neq l$, we denote by $X_{kl}$ the operator in
$\E^{\otimes m}$
which acts as $X$ in the product of $k$-th and $l$-th copies and as
1 in all other copies. That is,
$$
\aligned
&X=\sum_{r,s,t,u} a_{rstu} E_{rs}\otimes E_{tu}, \quad
a_{rstu}\in\C\qquad \Rightarrow\\
\Rightarrow \quad &X_{kl}=\sum_{r,s,t,u} a_{rstu} (E_{rs})_k\ (E_{tu})_l.
\endaligned\tag5
$$
Given formal variables $u_1,\dots,u_m$ we set for $k=1,\dots,m$
$$
T_k (u_k) :=\sum_{i,j=1}^N t_{ij} (u_k) \otimes (E_{ij})_k
\in \Y(N)[[u_1^{-1},\ldots,u_m^{-1}]]\otimes\End\E^{\otimes m}.
\tag 6
$$
We let $P$ denote the permutation operator in $\E\ot \E$:
$$
P:=\sum_{i,j}E_{ij}\ot E_{ji}.
$$
The defining
relations (1) can be rewritten now as the single {\it ternary relation} on
the $T$-matrix:
$$
R(u-v)T_1(u)T_2 (v) = T_2(v)T_1(u)R(u-v),   \tag 7
$$
where
$$
R(u)=R_{12}(u):=1-u^{-1}P_{12}
$$
is the {\it Yang} $R$-{\it matrix}.

\bigskip
\noindent {\bf 1.2.}
We shall regard the set of the matrix units $\{E_{ij}\}$ as a basis of the
Lie
algebra $\gl(N)$.
\proclaim
{\bf Theorem} The mapping
$$
\xi : t_{ij}(u)\mapsto \delta_{ij}+E_{ij}u^{-1} \tag1
$$
defines the algebra homomorphism
$$
\xi : \Y(N)\ra \U(\gl(N)).
$$
\endproclaim

\bigskip
\noindent
{\bf 1.3.}
Let $\Sym_m$ denote the symmetric group realized as the group of
permutations of the set $\{1,\ldots,m\}$ and let
$$
a_m=(m!)^{-1}\sum_{p\in \Sym_m}\sgn(p)\cdot p\in\C[\Sym_m]
$$
denote the normalized antisymmetrizer in the group ring. Consider the
natural
action of $\Sym_m$ in the tensor space $\E^{\otimes m}$ and denote by $A_m$
the
image of $a_m$.

\bigskip
\proclaim
{\bf Theorem} There exists a formal series
$$
\qdet T(u):=1+d_1u^{-1}+d_2u^{-2}+\cdots \in \Y(N)[[u^{-1}]] \tag 1
$$
such that the following identities hold:
$$
A_NT_1(u)\cdots T_N(u-N+1)=T_N(u-N+1)\cdots T_1(u)A_N=\qdet T(u)A_N.
\tag2
$$
\endproclaim

The series $\qdet T(u)$ is called the {\it quantum determinant}
of the matrix $T(u)$. Explicit formulae for the quantum determinant
can be easily derived from the identities (2).
Let $(i_1,\dots,i_N)$
be an arbitrary permutation of the indices $(1,\dots,N)$. Then
$$
\align
\qdet T(u)&=\sum_{p\in \Sym_N} \sgn(p)\ts t_{i_{p(1)},i_1}(u)\cdots
t_{i_{p(N)},i_N}(u-N+1)\tag3\\
&=\sum_{p\in \Sym_N} \sgn(p)\ts t_{i_1,i_{p(1)}}(u-N+1)\cdots
t_{i_N,i_{p(N)}}(u).\tag4
\endalign
$$

\bigskip
\proclaim
{\bf 1.4. Theorem} The coefficients $d_1,d_2,\dots$ of the quantum
determinant $\qdet T(u)$ are algebraically independent generators of
the center of the algebra $\Y(N)$.
\endproclaim

\bigskip
\noindent {\bf 1.5.}
Let $E$ denote the $N\times N$-matrix with the entries $E_{ij}$.
Then the image of the matrix $T(u)$ under the homomorphism (1.2.1) can
be written as $1+Eu^{-1}$. Denote by \newline $\qdet(1+Eu^{-1})$ the image
of
the quantum determinant $\qdet T(u)$ under this homomorphism. It follows
from
(1.3.3) that
$$
u(u-1)\dots(u-N+1)\ts\qdet(1+Eu^{-1})=
$$
\
$$
\det\left( \matrix
E_{11}+u&E_{12}&\hdots&E_{1N}\\
E_{21}&E_{22}+u-1&\hdots&E_{2N}\\
\vdots&\vdots&&\vdots\\
E_{N1}&E_{N2}&\hdots&E_{NN}+u-N+1
\endmatrix \right), \tag 1
$$
\bigskip
\noindent
where the determinant $\det A$ of a
noncommutative matrix $A=(a_{ij})_{i,j=1}^N$ is defined as
$$
\det A:=\sum_{p\in \Sym_N}\sgn(p)a_{p(1),1}\dots a_{p(N),N}.
$$
Denote the determinant (1) by $C(u)$. Then
$$
C(u)=u^N+z_1u^{N-1}+\dots+z_N, \qquad z_i\in \U(\gl(N)).\tag2
$$
Theorem 1.4 implies that all the coefficients $z_i$ belong to the
center of $\U(\gl(N))$. Let $\h$ denote the diagonal Cartan subalgebra
in $\gl(N)$. We shall identify the algebra $\U(\h)$ with the
algebra of polynomial functions on $\h^*$ and denote by $\lambda_i$ the
function which corresponds to $E_{ii}$. Then formula (1) easily implies
that the image $C_{\lambda}(u)$ of the polynomial $C(u)$ under the
Harish-Chandra homomorphism is given by the formula:
$$
C_{\lambda}(u)=\prod_{i=1}^N(u+\lambda_i-i+1). \tag3
$$
Therefore the images of the elements
$z_1,\dots,z_N$ are elementary symmetric
polynomials in the variables $l_1,\dots,l_N$, where $l_i=\lambda_i-i+1$.
So, we have the following theorem [HU, Appendix 1], [NT].

\bigskip
\proclaim
{\bf Theorem} The elements $z_1,\dots,z_N$ are algebraically
independent generators of the center of the algebra $\U(\gl(N))$.
\endproclaim

\bigskip
\noindent
{\bf 1.6.} Denote by $t$ the matrix transposition.
The following analogue of
the Cayley--Hamilton theorem holds for the matrices $E$ and $E^t$
[NT, Remark 2.5].

\proclaim
{\bf Theorem}
The matrices $E$ and $E^t$ satisfy the polynomial identities
$$
C(-E+N-1)=0
\qquad\text{and}\qquad
C(-E^t)=0. \tag 1
$$
\endproclaim

\Proof Let us define the matrix $\widehat T(u)=(\widehat t_{ij}(u))$ by the
following formula:
$$
A_NT_1(u)\cdots T_{N-1}(u-N+2)=A_N\widehat T_N(u). \tag2
$$
To rewrite this in terms of the matrix elements, for
$k_i\in\{1,\dots,N\}$, $i=1,\dots,N$ set
$$
\varepsilon(k_1,\dots,k_N)=\cases \sgn(p),&\text{if}\quad
(k_1,\dots,k_N)=(p(1),\dots,p(N)),\quad p\in S_N,\\
0,&\text{otherwise}.\endcases
$$
Then
$$
\varepsilon(i_1,\dots,i_N)\ts\widehat t_{i_Nj_N}(u)=
\sum_{a_1,\dots,a_{N-1}}\varepsilon(a_1,\dots,a_{N-1},j_N)\ts
t_{a_1i_1}(u)\cdots t_{a_{N-1}i_{N-1}}(u-N+2),\quad\tag3
$$
where $(i_1,\dots,i_N)$ is a permutation of the indices $(1,\dots,N)$
and $j_N\in\{1,\dots,N\}$.
One can easily see from (1.3.3) that $\widehat t_{ij}(u)$
equals $(-1)^{i+j}$ times the quantum determinant of the matrix obtained
from $T(u)$ by removing the $i$-th column and the $j$-th row, so the
elements $\widehat t_{ij}(u)$ are well-defined.

Formulae (1.3.2) and (2) imply that
$$
A_N\widehat T_N(u)T_N(u-N+1)=A_N\ts\qdet T(u),
$$
and hence,
$$
\qdet T(u)=\widehat T(u)T(u-N+1).\tag4
$$
This relation can be regarded as a quantum analogue of the
formula for
the inverse matrix of a numerical matrix.

It is not difficult to verify that the matrix $T^t(-u)$ satisfies
the ternary relation (1.1.7). In other words, the mapping
$$
\sigma:\ T(u)\ra T^t(-u)
$$
defines an automorphism of the algebra $\Y(N)$. Formulae (1.3.3) and
(1.3.4) imply that
$$
\sigma(\qdet T(u))=\qdet T(-u+N-1).
$$
Hence, applying $\sigma$, we obtain from (1.3.2) that
$$
A_NT^t_1(u-N+1)\cdots T^t_N(u)=A_N\qdet T(u).\tag5
$$
Let us define the matrix $\tilde T(u)$ by the formula
$$
A_NT^t_1(u-N+1)\cdots T^t_{N-1}(u-1)=A_N\tilde T_N(u).
$$
Comparing its matrix elements with those of the matrix
$\widehat T(u)$, we find from (1.3.3) and (1.3.4) that
$$
\tilde T(u)=\widehat T^t(u-1).
$$
So, identity (5) implies that
$$
\qdet T(u)=\widehat T^t(u-1)T^t(u).\tag6
$$

Let us apply the homomorphism $\xi$ (see Theorem 1.2) to both sides of the
identities (4) and (6). We have
$$
u\ts\xi(T(u))=u+E,
$$
and formula (3) implies that
$$
u(u-1)\cdots(u-N+2)\ts\xi(\widehat T(u))
$$
is a polynomial in $u$ with coefficients from $\U(\gl(N))\ot\End\E$,
which will be denoted by $\widehat C(u)$.
So, we obtain from identities (4)
and (6) that
$$
C(u)=\widehat C(u)(u+E-N+1)
$$
and
$$
C(u)=\widehat C^t(u-1)(u+E^t)
$$
which proves the theorem.

\bigskip
\noindent {\bf 1.7.} Let
$L(\lambda)$, $\lambda=(\lambda_1,\dots,\lambda_N)\in\h^*$ denote
the highest weight
representation of
the Lie algebra $\gl(N)$. It is clear that
the image of the polynomial $C(u)$ in
$L(\lambda)$ coincides with $C_{\lambda}(u)$.
So, Theorem 1.6 implies the following corollary [NT] (cf. [BG], [Gr]).
\proclaim
{\bf Corollary} The images of the matrices $E$ and $E^t$ in $L(\lambda)$
satisfy the identities
$$
\prod_{i=1}^N(E-\lambda_i+i-N)=0\quad\text{and}\quad
\prod_{i=1}^N(E^t-\lambda_i+i-1)=0.
$$
\endproclaim

\newpage
\heading
{\bf 2. The twisted Yangian $\Y^{\pm}(N)$ and the Sklyanin determinant}
\endheading
\

In this section following Olshanski\u\i\ [O2], [MNO] we define the algebra
$\Y^{\pm}(N)$ and outline the construction of the Sklyanin determinant.

\bigskip
\noindent {\bf 2.1.}
As before, we will denote by $\E$
the vector space $\C^N$ and let $\{e_i\}$
be its canonical basis. From now on it will be convenient to parametrize
the basis vectors by the numbers $i=-n,-n+1,\ldots,n-1,n$, where $n:=[N/2]$
and $i=0$ is skipped when $N$ is even.

Let us equip $\E$ with a nondegenerate bilinear form
which may be either symmetric or alternating:
$$
<e_i,e_j>_+=\delta_{i,-j},\qquad <e_i,e_j>_-=\sgn(i)\delta_{i,-j}.
$$
Clearly, the alternating case may occur only when $N$ is even.

Both of the cases, symmetric and alternating, will be considered
simultaneously unless stated otherwise. It will be convenient to use
the symbol $\theta_{ij}$ which is defined as follows:
\medskip
$$
\theta_{ij}:=\cases 1,\quad&\text{in the symmetric case};\\
\sgn(i)\sgn(j),\quad&\text{in the alternating case}.\endcases
$$
\bigskip

Whenever the double sign $\pm{}$ or $\mp{}$ occurs,
the upper sign corresponds to the symmetric case and the lower sign to
the alternating one.

By $X\mapsto X^t$ we will denote the
transposition with respect to the form $<\cdot,\cdot>_{\pm}$, which is
an antiautomorphism of the algebra $\End\E$ such that
$$
(E_{ij})^t=\theta_{ij}E_{-j,-i}.
$$
Given tensor product $\End\E^{\ot m}$
we denote by $t_k$ the partial
transposition corresponding to the $k$-th copy of $\E$.

\bigskip
\noindent {\bf 2.2.}
The {\it twisted Yangian} $\Y^{\pm}(N)$ is defined as a subalgebra in
the Yangian $\Y(N)$ in the following way. Let us introduce the $S$-{\it
matrix} by setting
$$
S(u):=T(u)T^t(-u). \tag 1
$$
Then
$$
S(u)=\sum_{i,j} s_{ij}(u)\ot E_{ij}
$$
with
$$
s_{ij}(u)=\delta_{ij}+s_{ij}^{(1)}u^{-1}+s_{ij}^{(2)}u^{-2}+\cdots, \tag 2
$$
and $\Y^{\pm}(N)$ is the subalgebra of $\Y(N)$ generated by
the elements $s_{ij}^{(1)},s_{ij}^{(2)},\dots$, where $-n\leq i,j\leq n$.

Let us introduce the following element of the algebra $\End\E^{\ot2}$:
$$
Q:=P^{t_1}=P^{t_2}=
\sum_{i,j}\theta_{ij}E_{-j,-i}\ot E_{ji},
$$
and set
$$
R'(u)=1-u^{-1}Q.
$$

\bigskip
\proclaim
{\bf 2.3. Theorem} The $S$-matrix satisfies the relations:
$$
R(u-v)S_1(u)R'(-u-v)S_2(v)=S_2(v)R'(-u-v)S_1(u)R(u-v), \tag 1
$$
$$
S^t(-u)=S(u)\pm {S(u)-S(-u)\over 2u}.\tag 2
$$
\endproclaim

Relations (1) and (2) are called the {\it quaternary relation} and
the {\it symmetry relation} respectively. They can be rewritten in
terms of the generating series $s_{ij}(u)$ as follows:
$$
\align
[s_{ij}(u),s_{kl}(v)]=&{1\over
u-v}(s_{kj}(u)s_{il}(v)-s_{kj}(v)s_{il}(u))\\
-&{1\over u+v}(\theta_{k,-j}s_{i,-k}(u)s_{-j,l}(v)-
\theta_{i,-l}s_{k,-i}(v)s_{-l,j}(u))\\
+&{1\over u^2-v^2}(\theta_{i,-j}s_{k,-i}(u)s_{-j,l}(v)-
\theta_{i,-j}s_{k,-i}(v)s_{-j,l}(u))
\endalign
$$
and
$$
\theta_{ij}s_{-j,-i}(-u)=s_{ij}(u)\pm {s_{ij}(u)-s_{ij}(-u)\over 2u}. \tag
3
$$

\bigskip
\proclaim
{\bf 2.4. Theorem} The quaternary relation
and the symmetry
relation are precisely the defining relations for the
algebra $\Y^{\pm}(N)$.
\endproclaim

\bigskip
\noindent {\bf 2.5.}
Define the matrix $F$ by the formula
$$
F:=E-E^t.
$$
Its matrix elements have the form
$$
F_{ij}=E_{ij}-\theta_{ij}E_{-j,-i},\qquad -n\leq i,j\leq n.
$$
Denote by $\g(n)$ the Lie subalgebra of $\gl(N)$ spanned
by the elements $F_{ij}$.
Then $\g(n)$ is isomorphic to $\oa(2n)$ or
$\spa(2n)$ if $N=2n$, and  $\oa(2n+1)$ if $N=2n+1$. The
generators $F_{ij}$ satisfy the following relations:
$$
[F_{ij},F_{kl}]=\delta_{kj}F_{il}-\delta_{il}F_{kj}-
\theta_{ij}(\delta_{k,-i}F_{-j,l}-\delta_{-j,l}F_{k,-i}),
$$
and
$$
F_{-j,-i}=-\theta_{ij}F_{ij}.\tag 1
$$

\proclaim
{\bf Theorem} The mapping
$$
\xi: s_{ij}(u)\mapsto \delta_{ij}+F_{ij}(u\pm {1\over 2})^{-1} \tag 2
$$
defines the algebra homomorphism
$$
\xi: \Y^{\pm}(N)\ra \U(\g(n)).
$$
\endproclaim

\bigskip
\noindent {\bf 2.6.}
We shall keep using the notation (1.1.4)--(1.1.6).

\proclaim
{\bf Theorem} There exists a formal series
$$
\sdet S(u)\in \Y^{\pm}(N)[[u^{-1}]]
$$
such that the following identities hold
$$
\aligned
A_NS_1(u)\tilde R_1S_2(u-1)&\tilde R_2\cdots
S_N(u-N+1)=\\
&S_N(u-N+1)\cdots \tilde R_2S_2(u-1)
\tilde R_1S_1(u)A_N =\sdet S(u)A_N,
\endaligned\tag 1
$$
where
$$
\tilde R_i=R'_{i,i+1}\cdots R'_{i,N},\qquad R'_{ij}=R'_{ij}(-2u+i+j-2).
$$
\endproclaim

The series $\sdet S(u)$ is called
the {\it Sklyanin determinant} of the matrix $S(u)$.

\bigskip
\noindent {\bf 2.7.}
The coefficients of $\sdet S(u)$ can be expressed in terms of those
of the quantum determinant $\qdet T(u)$ in the following way.

\proclaim
{\bf Theorem}
$$
\sdet S(u)=\gamma_N(u)\ts\qdet T(u)\ts\qdet T(-u+N-1),
$$
where $\gamma_N(u)\equiv1$ for $\Y^+(N)$ and $\dsize\gamma_N(u)=
\frac{2u+1}{2u-N+1}$ for $\Y^-(N)$.
\endproclaim

In particular, $\sdet S(u)$ satisfies the relation
$$
\gamma_N^{-1}(u)\ts\sdet S(u)=\gamma_N^{-1}(-u+N-1)\ts\sdet S(-u+N-1).
\tag 1
$$

\bigskip
\proclaim
{\bf 2.8. Theorem} We have
$$
\sdet S(u+{N\over 2}-{1\over 2})=1+c_2u^{-2}+c_4u^{-4}+\cdots
$$
in the orthogonal case, and
$$
\sdet S(u+n-{1\over 2})=(1+nu^{-1})(1+c_2u^{-2}+c_4u^{-4}+\cdots)
$$
in the symplectic case. Moreover, the elements $c_2,c_4,\dots$ are
algebraically independent generators of the center of the algebra
$\Y^{\pm}(N)$.
\endproclaim

\newpage
\heading
{\bf 3. Formulae for $\sdet S(u)$}
\endheading
\

Using formula (2.6.1) one can express the Sklyanin determinant
$\sdet S(u)$ in terms of the series $s_{ij}(u)$ directly. However,
because of the intermediate factors $R'_{ij}$,
this expression turns out to be too
complicated
to be very useful.
Our aim now is to obtain `short' formulae
for $\sdet S(u)$ analogous to (1.3.3) and (1.3.4) for the
quantum determinant (Theorem 3.6).
The basic idea is to eliminate the factors $R'_{ij}$ by
using the symmetry relation (2.3.2). We maintain the
notation of Section 2.

\bigskip
\noindent {\bf 3.1.}
Let $(a_1,\dots,a_m),\ m\geq 3$ be an ordered set. Define
transformations of ordered pairs $(a_k,a_l),\ k\ne l$
by the following rule:
$$
\aligned
(a_k,a_l)&\ra (a_l,a_k),\quad k,l<m,\\
(a_k,a_m)&\ra (a_{m-1},a_k), \quad k<m-1,\\
(a_m,a_k)&\ra (a_k,a_{m-1}), \quad k<m-1,\\
(a_{m-1},a_m)&\ra (a_{m-1},a_{m-2}),\\
(a_m,a_{m-1})&\ra (a_{m-1},a_{m-2}).
\endaligned \tag 1
$$
Furthermore, given an ordered set $(a_1,a_2)$ we define the
transformation which passes each ordered pair $(a_1,a_2)$ and $(a_2,a_1)$
into the element $a_1$.
Using these transformations define a map
$$
\Sym_N\ra \Sym_{N-1},\qquad p\ra p' \tag 2
$$
in the following way. (We regard $\Sym_{N-1}$ as a
natural subgroup in $\Sym_N$). Consider the ordered set
$(1,2,\dots,N)$ and take as $(p'(1),p'(N-1))$
the image of the ordered pair $(p(1),p(N))$
under the transformation (1). Then we consider the
ordered set obtained from $(1,2,\dots,N)$ by removing the
numbers $p(1)$ and $p(N)$ and repeat the previous step
for the ordered pair $(p(2),p(N-1))$ to obtain
the pair $(p'(2),p'(N-2))$ etc.
If $N=2n+1$, repeating this procedure,
after the $n$-th step we get a permutation $(p'(1),\dots,p'(N-1))$
of the indices $(1,\dots,N-1)$.
Similarly, if $N=2n$
the procedure ends by using for the $n$-th step the
described above transformation
for a two element set.

The projection (2) will be used in our formulae for
$\sdet S(u)$, so we give here a brief
description of its combinatorial properties.

Let us consider a graph $\Gamma_N$ whose vertices are
identified with the elements of the symmetric group
$\Sym_N$ and two permutations $p=(p(1),\dots,p(N))$
and $q=(q(1),\dots,q(N))$ are connected by an edge if $q$ can be
obtained from $p$ by interchanging two indices
$p(k)$ and $p(l)$, provided that
either
\smallskip

(i) $p(k)$ and $p(l)$ are the two maximal elements of the set
$\{p(i),p(i+1),\dots,p(N-i+1)\}$ for some $1\leq i\leq n$; or
\smallskip

(ii) $k<l<N-k+1$ and $p(k),p(l),p(N-k+1)$ are the three maximal
elements of the set $\{p(k),p(k+1),\dots,p(N-k+1)\}$.
\smallskip

Note that if $p$ and $q$ are connected by an edge in $\Gamma_N$ then
$p$ and $q$ are compatible with respect to the {\it Bruhat order}
on $\Sym_N$.

We shall need {\it signless Stirling numbers of the first kind}
$c\thinspace(m,k)$
which are defined by the formula
$$
\sum_{k=1}^m c\thinspace(m,k)x^k=x(x+1)\cdots (x+m-1).
$$
The following proposition can be easily derived from the definition of the
map
$p\mapsto p'$ with the use of some properties of the numbers
$c\thinspace(m,k)$ (see, e.g. Stanley [S]).

\bigskip
\proclaim
{\bf Proposition} {\rm (i)} The graph $\Gamma_N$ has $(N-1)!$
connected components and each of them is isomorphic (as a graph) to
the $k$-dimensional cube for some $1\leq k\leq N-1$. Moreover,
given $k$ the number of components, isomorphic to
the $k$-dimensional cube, equals
$c\thinspace(N-1,k)$.

{\rm (ii)} Each connected component of the graph $\Gamma_N$ contains
a unique vertex\newline$p=(p(1),\dots,p(N))$ such that $p(n+1)=N$, and all
the vertices of this component have the same image
$p'=(p(N),\cdots,p(n+2),p(n),\cdots,p(1))$ under the projection
{\rm (2)}.
\endproclaim

\bigskip
\noindent
{\bf 3.2. Examples.} Let us represent elements of the symmetric group
in the form
$$
p=(p(1),p(2),\dots,p(N)).
$$
Then we have:
$$
\alignat2
&N=2&&N=3\\
&p\ \ \ts
(1,2)\ \ (2,1)\qquad\qquad &&p\ \ \ts(1,2,3)\ \ (1,3,2)\
\ (2,1,3)\ \ (2,3,1)\ \ (3,1,2)\ \ (3,2,1)\\
&p'\ \ \ (1)\quad\ \ (1) &&p'\ \ \ (2,1)\quad\ \ (2,1)
\quad\ \ (2,1)\quad\ \ (1,2)\quad\ \ (2,1)\quad\ \ (1,2)
\endalignat
$$
$$
\align
&N=4\\
&p\ \ \ts(1,2,3,4)\ \ (1,2,4,3)\ \ (1,3,2,4)\ \ (1,3,4,2)\ \ (1,4,2,3)
\ \ (1,4,3,2)\ \ (2,1,3,4)\ \ (2,1,4,3)\\
&p'\ \ \ (3,2,1)\quad\ \ (3,2,1)\quad\ \ (3,2,1)\quad\ \ (2,3,1)\quad\ \
(3,2,1)\quad\ \ (2,3,1)\quad\ \ (3,1,2)\quad\ \ (3,1,2)
\endalign
$$
$$
\align
&p\ \ \ts(2,3,1,4)\ \ (2,3,4,1)\ \ (2,4,1,3)\ \ (2,4,3,1)\ \ (3,1,2,4)
\ \ (3,1,4,2)\ \ (3,2,1,4)\ \ (3,2,4,1)\\
&p'\ \ \ (3,1,2)\quad\ \ (1,3,2)\quad\ \ (3,1,2)\quad\ \ (1,3,2)\quad\ \
(3,1,2)\quad\ \ (2,1,3)\quad\ \ (3,1,2)\quad\ \ (1,2,3)
\endalign
$$
$$
\align
&p\ \ \ts(3,4,1,2)\ \ (3,4,2,1)\ \ (4,1,2,3)\ \ (4,1,3,2)\ \ (4,2,1,3)
\ \ (4,2,3,1)\ \ (4,3,1,2)\ \ (4,3,2,1)\\
&p'\ \ \ (2,1,3)\quad\ \ (1,2,3)\quad\ \ (3,1,2)\quad\ \ (2,1,3)\quad\ \
(3,1,2)\quad\ \ (1,2,3)\quad\ \ (2,1,3)\quad\ \ (1,2,3)
\endalign
$$

\bigskip
\noindent {\bf 3.3.} For $i=1,\dots,m$ denote by $A_m^{(i)}$
the normalized antisymmetrizer in the
tensor space $\E^{\ot m}$, which corresponds to
the subgroup of $\Sym_m$ consisting of the permutations which
preserve each of the first $i-1$ indices. In particular, $A_m^{(1)}$
coincides
with $A_m$.

\proclaim
{\bf Proposition} For any $m\geq 2$ one has the equalities
$$
\align
A_m^{(2)}R'_{12}\cdots R'_{1m}&=R'_{1m}\cdots R'_{12}A_m^{(2)}\tag1\\
&=(1+\frac1{2u-1}(Q_{12}+\cdots+Q_{1m}))A_m^{(2)}, \tag 2
\endalign
$$
where, as before, $R'_{1i}=R'_{1i}(-2u+i-1)$.
\endproclaim

\Proof By definition of $R'_{1i}(u)$ we can write
$$
A_m^{(2)}R'_{12}\cdots R'_{1m}=
A_m^{(2)}(1+\frac{Q_{12}}{2u-1})\cdots (1+\frac{Q_{1m}}{2u-m+1}).\tag3
$$
Let us prove that for $1<i_1<\cdots<i_k$
$$
A_m^{(2)}Q_{1i_1}\cdots Q_{1i_k}=(-1)^{k-1}A_m^{(2)}Q_{1i_k}.\tag4
$$
Indeed, it is clear that
$$
P_{1i_k}P_{1i_{k-1}}\cdots P_{1i_1}=P_{i_ki_{k-1}}\cdots
P_{i_ki_1}P_{1i_k}.
$$
Hence,
$$
A_m^{(2)}P_{1i_k}\cdots P_{1i_1}=(-1)^{k-1}A_m^{(2)}P_{1i_k}.
$$
Applying the transposition $t_1$ we get (4). Formula (4) allows us
to rewrite (3) in the form
$$
A_m^{(2)}(1+a_2(u)Q_{12}+\cdots+a_m(u)Q_{1m}),
$$
where
$$
a_i(u)=\sum\frac{(-1)^k}{(2u-i_1+1)\cdots(2u-i_k+1)(2u-i+1)}
$$
and the sum is taken over the sets of the indices $(i_1,\dots,i_k)$ such
that $1<i_1<\cdots<i_k<i$, $k\geq 0$. It is easy to see that
$$
a_i(u)=\frac1{2u-i+1}\prod_{j=2}^{i-1}(1-\frac1{2u-j+1})=\frac1{2u-1}.
$$
Observe that
$$
A_m^{(2)}(P_{12}+\cdots+P_{1m})=(P_{12}+\cdots+P_{1m})A_m^{(2)}.
$$
Applying the transposition $t_1$ we obtain that
$$
A_m^{(2)}(Q_{12}+\cdots+Q_{1m})=(Q_{12}+\cdots+Q_{1m})A_m^{(2)},\tag5
$$
which proves the equality
$$
A_m^{(2)}R'_{12}\cdots R'_{1m}
=(1+\frac1{2u-1}(Q_{12}+\cdots+Q_{1m}))A_m^{(2)}.\tag6
$$
Similar arguments show that $R'_{1m}\cdots R'_{12}A_m^{(2)}$ coincides with
the right hand side of (6), which completes the proof.

\bigskip
\noindent {\bf 3.4.}
For $m\leq N$ define the formal series
$$
\si_{i_1,\dots,i_m}^{a_1,\dots,a_m}(u),\ \
\tsi_{i_1,\dots,i_{m-1},j}^{a_1,\dots,a_m}(u)\in\Y^{\pm}(N)[[u^{-1}]]
$$
where $i_1,\dots,i_m,a_1,\dots,a_m,j\in\{-n,-n+1,\dots,n\}$, as follows:
$$
\align
A_mS_1(u)R'_{12}\cdots R'_{1m}S_2&(u-1)
R'_{23}\cdots R'_{2m}S_3(u-2)\\
\cdots &S_{m-1}(u-m+2)R'_{m-1,m}S_m(u-m+1)
(e_{i_1}\ot\cdots\ot e_{i_m})
\endalign
$$
$$
=\sum_{a_1,\dots,a_m}\si_{i_1,\dots,i_m}^{a_1,\dots,a_m}(u)
(e_{a_1}\ot\cdots\ot e_{a_m})
\tag 1
$$
and
$$
\align
A_mS_1(u)R'_{12}\cdots R'_{1m}S_2&(u-1)
R'_{23}\cdots R'_{2m}S_3(u-2)\\
\cdots &S_{m-1}(u-m+2)R'_{m-1,m}
(e_{i_1}\ot\cdots\ot e_{i_{m-1}}\ot e_j)
\endalign
$$
$$
=\sum_{a_1,\dots,a_m}\tsi_{i_1,\dots,i_{m-1},j}^{a_1,\dots,a_m}(u)
(e_{a_1}\ot\cdots\ot e_{a_m}).
\tag 2
$$
In particular, $\si_{i_1}^{a_1}(u)=s_{a_1i_1}(u)$ and if $(i_1,\dots,i_N)$
is a permutation of the indices \newline
$(-n,-n+1,\dots,n)$ then by (2.6.1)
$$
N!\ts\si_{i_1,\dots,i_N}^{i_{q(1)},\dots,i_{q(N)}}(u)=
\sgn(q)\ts\sdet S(u),\qquad q\in \Sym_N.\tag 3
$$
It follows immediately from (1) and (2) that
$$
\si_{i_1,\dots,i_m}^{a_1,\dots,a_m}(u)=\sum_j
\tsi_{i_1,\dots,i_{m-1},j}^{a_1,\dots,a_m}(u)\ts s_{ji_m}(u-m+1)\tag 4
$$
and for any $q\in\Sym_m$
$$
\si_{i_1,\dots,i_m}^{a_{q(1)},\dots,a_{q(m)}}(u)=\sgn(q)\ts
\si_{i_1,\dots,i_m}^{a_1,\dots,a_m}(u) \tag 5
$$
and
$$
\tsi_{i_1,\dots,i_{m-1},j}^{a_{q(1)},\dots,a_{q(m)}}(u)=\sgn(q)\ts
\tsi_{i_1,\dots,i_{m-1},j}^{a_1,\dots,a_m}(u).\tag 6
$$
Furthermore, the following proposition holds.

\proclaim
{\bf Proposition} For any $q\in\Sym_m$ and $r\in\Sym_{m-1}$
$$
\si_{i_{q(1)},\dots,i_{q(m)}}^{a_1,\dots,a_m}(u)=\sgn(q)\ts
\si_{i_1,\dots,i_m}^{a_1,\dots,a_m}(u)\tag 7
$$
and
$$
\tsi_{i_{r(1)},\dots,i_{r(m-1)},j}^{a_1,\dots,a_m}(u)=\sgn(r)\ts
\tsi_{i_1,\dots,i_{m-1},j}^{a_1,\dots,a_m}(u).\tag 8
$$
\endproclaim

\Proof We use the following identity
$$
\aligned
A_mS_1(u)R'_{12}\cdots R'_{1m}S_2(u-1)
R'_{23}&\cdots R'_{2m}S_3(u-2)\\
\cdots &S_{m-1}(u-m+2)R'_{m-1,m}S_m(u-m+1)\\
=S_m(u-m+1)R'_{m-1,m}S_{m-1}&(u-m+2)\\
\cdots S_3(u-2)&R'_{2m}\cdots R'_{23}S_2(u-1)
R'_{1m}\cdots R'_{12}S_1(u)A_m;
\endaligned\tag 9
$$
see [O2, Lemmas 1.5, 2.5]. Clearly, (9) implies (7).
To prove (8) we note that $R'_{ij}$ is permutable with
$R'_{kl}$ and $S_k(u)$, provided that the indices $i,j,k,l$ are distinct.
Hence, we can rewrite the operator in the left hand side of (2) as follows
$$
\aligned
A_mS_1(u)R'_{12}\cdots R'_{1,m-1}S_2(u-1)
R'_{23}&\cdots R'_{2,m-1}S_3(u-2)\\
\cdots &S_{m-1}(u-m+2)R'_{1m}\cdots R'_{m-1,m}.
\endaligned\tag 10
$$
Clearly, $A_m=A_mA_{m-1}$. Therefore, applying (9) with $m$ replaced
by $m-1$, we bring (10) to the form:
$$
\align
A_mS_{m-1}(u-m+2)\cdots S_3(u-2)R'_{2,m-1}\cdots R'_{23}&\\
\cdot S_2(u-1)R'_{1,m-1}\cdots R'_{12}&S_1(u)A_{m-1}R'_{1m}\cdots
R'_{m-1,m}.
\endalign
$$
Finally, using the same arguments as in the proof of Proposition 3.3,
one can obtain the following analogue of identity (3.3.1)
$$
A_{m-1}R'_{1m}\cdots R'_{m-1,m}=
R'_{m-1,m}\cdots R'_{1m}A_{m-1}
$$
which implies (8). The proposition is proved.

\bigskip
\proclaim
{\bf 3.5. Proposition} Suppose that $-i_1\in\{a_1,\dots,a_m\}$
and $j\notin\{-i_2,\dots,-i_{m-1}\}$. Then
$$
\align
&\tsi_{i_1,\dots,i_{m-1},j}^{a_1,\dots,a_m}(u)=0\quad\text{if}\quad
j\notin\{a_1,\dots,a_m\};\quad\text{and}\tag1\\
&\tsi_{i_1,\dots,i_{m-1},j}^{a_1,\dots,a_m}(u)=
\frac1{m(m-1)}\frac{2u+1}{2u\pm1}\sum_{k=1}^{m-1}(-1)^{k-1}s_{a_ki_1}^t(-u)
\ts\si_{i_2,\dots,i_{m-1}}^{a_1,\dots,\widehat{a_k},\dots,a_{m-1}}(u-1),
\tag2
\endalign
$$
if $j=a_m$; here the sign $\ \widehat{}\ $ indicates that the corresponding

index should be missed.
\endproclaim

\Proof Let us represent the left hand side of (3.4.2)
in the form
$$
\aligned
A_mS_1(u)R'_{12}\cdots R'_{1,m-1}R'_{1m}&S_2(u-1)
R'_{23}\cdots R'_{2,m-1}S_3(u-2)\\
\cdots &S_{m-1}(u-m+2)R'_{2m}\cdots R'_{m-1,m}
(e_{i_1}\ot\cdots\ot e_{i_{m-1}}\ot e_j)
\endaligned\tag 3
$$
and transform it in the following way. First we note that
the assumption \newline
$j\notin\{-i_2,\dots,-i_{m-1}\}$
implies that
$$
R'_{2m}\cdots R'_{m-1,m}(e_{i_1}\ot\cdots\ot e_{i_{m-1}}\ot e_j)=
e_{i_1}\ot\cdots\ot e_{i_{m-1}}\ot e_j.
$$
Further we write $A_m=A_mA_m^{(2)}$ and
permute $A_m^{(2)}$
with $S_1(u)$. Then
applying Proposition 3.3 we replace $A_m^{(2)}R'_{12}\cdots R'_{1m}$
with
$$
(1+\frac1{2u-1}(Q_{12}+\cdots+Q_{1m}))A_m^{(2)}.
$$
Now we write
$A_m^{(2)}=A_m^{(2)}A_{m-1}^{(2)}$ and using (3.4.1) bring
(3) to the form
$$
\align
A_mS_1(u)(1+\frac1{2u-1}(Q_{12}+\cdots+Q_{1m}))&\\
\cdot A_m^{(2)}\sum_{b_2,\dots,b_{m-1}}
\si_{i_2,\dots,i_{m-1}}^{b_2,\dots,b_{m-1}}&(u-1)(e_{i_1}\ot
e_{b_2}\ot\dots
\ot e_{b_{m-1}}\ot e_j).\tag 4
\endalign
$$
Finally, applying (3.3.5) we permute
$A_m^{(2)}$ with
$Q_{12}+\cdots+Q_{1m}$ and $S_1(u)$, so that (4) looks now as
$$
A_mS_1(u)(1+\frac1{2u-1}(Q_{12}+\cdots+Q_{1m}))
\sum_{b_2,\dots,b_{m-1}}
\si_{i_2,\dots,i_{m-1}}^{b_2,\dots,b_{m-1}}(u-1)(e_{i_1}\ot e_{b_2}\ot\dots
\ot e_{b_{m-1}}\ot e_j).\tag 5
$$
If the set $\{a_1,\dots,a_m\}$ does not contain $j$ then,
in particular, $j\ne -i_1$ and hence
$$
Q_{1m}(e_{i_1}\ot e_{b_2}\ot\dots
\ot e_{b_{m-1}}\ot e_j)=0.\tag 6
$$
So, decomposing (5)
into a linear combination of the basis vectors, we see that
this decomposition only contains vectors of the form
$e_{a_1}\ot\dots\ot e_{a_m}$ with $j\in\{a_1,\dots,a_m\}$
which proves (1).

Let us suppose now that $j=a_m$ and $a_m\ne -i_1$. Then relation (6)
will still hold. Using the definition of $Q$ we can rewrite (5) as
follows:
$$
\align
A_m\sum_{b_1,\dots,b_{m-1}} \bigl(s_{b_1i_1}(u)\ts
\si_{i_2,\dots,i_{m-1}}^{b_2,\dots,b_{m-1}}(u-1)
+\frac1{2u-1}\theta_{-b_2,i_1}s_{b_1,-b_2}(u)\ts
\si_{i_2,\dots,i_{m-1}}^{-i_1,b_3,\dots,b_{m-1}}(u-1)&\\
+\cdots+\frac1{2u-1}\theta_{-b_{m-1},i_1}s_{b_1,-b_{m-1}}(u)\ts
\si_{i_2,\dots,i_{m-1}}^{b_2,\dots,b_{m-2},-i_1}(u-1)\bigr)
(e_{b_1}\ot\dots
\ot e_{b_{m-1}}\ot e_j)&.
\endalign
$$
We are interested in the coefficient of $e_{a_1}\ot\dots\ot e_{a_{m-1}}\ot
e_j$ in this expression. Relations (3.4.5) and (3.4.6) imply that without
loss of generality we may assume that $a_1=-i_1$. Then the required
coefficient is
$$
\aligned
\frac1{m!}&\sum_{q\in\Sym_{m-1}}\sgn(q)s_{a_{q(1)},i_1}(u)\ts
\si_{i_2,\dots,i_{m-1}}^{a_{q(2)},\dots,a_{q(m-1)}}(u-1)\\
+\frac1{m!}\frac1{2u-1}&\sum_{q\in\Sym_{m-1}}\sgn(q)
\theta_{-a_{q(2)},i_1}s_{a_{q(1)},-a_{q(2)}}(u)\ts
\si_{i_2,\dots,i_{m-1}}^{a_1,a_{q(3)},\dots,a_{q(m-1)}}(u-1)+\dots\\
+\frac1{m!}\frac1{2u-1}&\sum_{q\in\Sym_{m-1}}\sgn(q)
\theta_{-a_{q(m-1)},i_1}s_{a_{q(1)},-a_{q(m-1)}}(u)\ts
\si_{i_2,\dots,i_{m-1}}^{a_{q(2)},\dots,a_{q(m-2)},a_1}(u-1).
\endaligned\tag 7
$$
Using again (3.4.5) we find that for any $k=1,\dots,m-1$
$$
\sum_{q\in\Sym_{m-1},\ q(1)=k}\sgn(q)\ts
\si_{i_2,\dots,i_{m-1}}^{a_{q(2)},\dots,a_{q(m-1)}}(u-1)=
(m-2)!\ts(-1)^{k-1}
\si_{i_2,\dots,i_{m-1}}^{a_1,\dots,\widehat{a_k},\dots a_{m-1}}(u-1)\tag 8
$$
and for any $k,l=2,\dots,m-1$
$$
\sum\sgn(q)\ts
\si_{i_2,\dots,i_{m-1}}^{a_{q(2)},\dots,a_{q(l-1)},
a_1,a_{q(l+1)},\dots,a_{q(m-1)}}(u-1)=
(m-3)!\ts(-1)^{k}
\si_{i_2,\dots,i_{m-1}}^{a_1,\dots,\widehat{a_k},\dots a_{m-1}}(u-1);
$$
here the sum is taken over the permutations $q\in\Sym_{m-1}$ such that
$q(1)=1$, $q(l)=k$. Taking the sum in the left hand side over those
permutations with $q(1)=k$, $q(l)=1$ we clearly get the same formula
with the minus sign on the right.
This enables us
to rewrite (7) in the form
$$
\align
\frac1{m(m-1)}&\bigl(s_{-i_1,i_1}(u)\ts
\si_{i_2,\dots,i_{m-1}}^{a_2,\dots,a_{m-1}}(u-1)\\
+\sum_{k=2}^{m-1}&(-1)^{k-1}(\frac{2u}{2u-1}s_{a_ki_1}(u)-\frac1{2u-1}
\theta_{-a_k,i_1}s_{-i_1,-a_k}(u))\ts
\si_{i_2,\dots,i_{m-1}}^{a_1,\dots,\widehat{a_k},\dots a_{m-1}}(u-1)\bigr).
\endalign
$$
Note that $\theta_{-a_k,i_1}=\pm\theta_{a_k,i_1}$.
Now the symmetry relation (2.3.3) gives
$$
s_{-i_1,i_1}(u)=\frac{2u+1}{2u\pm1}s_{-i_1,i_1}^t(-u)
$$
and
$$
\frac{2u}{2u-1}s_{a_ki_1}(u)\mp\frac1{2u-1}
\theta_{a_k,i_1}s_{-i_1,-a_k}(u)=\frac{2u+1}{2u\pm1}s_{a_ki_1}^t(-u)
$$
which proves (2) in the case under consideration.

Suppose now that $j=a_m=-i_1$. Comparing (4) and (5) we conclude that
the summation in (5) may be taken over the indices $b_k$ such that
$b_k\ne -i_1$ for $k=2,\dots,m-1$. Then
$$
Q_{1k}(e_{i_1}\ot e_{b_2}\ot\dots
\ot e_{b_{m-1}}\ot e_j)=0
$$
for $k=2,\dots,m-1$, and so, expression (5) takes the form
$$
A_mS_1(u)(1+\frac1{2u-1}Q_{1m})
\sum_{b_2,\dots,b_{m-1}}
\si_{i_2,\dots,i_{m-1}}^{b_2,\dots,b_{m-1}}(u-1)(e_{i_1}\ot e_{b_2}\ot\dots
\ot e_{b_{m-1}}\ot e_j),
$$
which equals
$$
\align
A_m\bigl(\sum_{b_1,\dots,b_{m-1}} s_{b_1i_1}(u)\ts
\si_{i_2,\dots,i_{m-1}}^{b_2,\dots,b_{m-1}}(u-1)
(e_{b_1}\ot\dots
\ot e_{b_{m-1}}\ot e_{-i_1})\qquad\qquad\qquad\qquad&\\
+\frac1{2u-1}\sum_{b_1,\dots,b_m}\theta_{-b_m,i_1}s_{b_1,-b_m}(u)\ts
\si_{i_2,\dots,i_{m-1}}^{b_2,\dots,b_{m-1}}(u-1)
(e_{b_1}\ot\dots
\ot e_{b_m})\bigr)&,
\endalign
$$
where the indices $b_1$ and $b_m$ run through the set $\{-n,\dots,n\}$.
The coefficient of \newline
$e_{a_1}\ot\dots
\ot e_{a_{m-1}}\ot e_{-i_1}$ in this expression is
$$
\align
\frac1{m!}\frac{2u}{2u-1}&\sum_{q\in\Sym_{m-1}}\sgn(q)s_{a_{q(1)},i_1}(u)\ts
\si_{i_2,\dots,i_{m-1}}^{a_{q(2)},\dots,a_{q(m-1)}}(u-1)\\
-\frac1{m!}\frac1{2u-1}&\sum_{q\in\Sym_{m-1}}\sgn(q)
\theta_{-a_{q(1)},i_1}s_{-i_1,-a_{q(1)}}(u)\ts
\si_{i_2,\dots,i_{m-1}}^{a_{q(2)},\dots,a_{q(m-1)}}(u-1).
\endalign
$$
Applying (8) and the symmetry relation (2.3.3) we bring this again to the
form
(2). The proposition is proved.

\bigskip
\proclaim
{\bf 3.6. Theorem} Let $(i_1,\dots,i_N)$ be an arbitrary permutation of
the indices \newline
$(-n,-n+1,\dots,n)$. Then
$$
\aligned
\sdet S(u)=(-1)^n\gamma_N(u)\sum_{p\in\Sym_N}\sgn(pp')
s^t_{-i_{p(1)},i_{p'(1)}}(-u)\cdots
s^t_{-i_{p(n)},i_{p'(n)}}(-u+n-1)&\\
\cdot s_{-i_{p(n+1)},i_{p'(n+1)}}(u-n)\cdots
s_{-i_{p(N)},i_{p'(N)}}(u-N+1)&
\endaligned\tag1
$$
and also
$$
\aligned
=(-1)^n\gamma_N(u)\sum_{p\in\Sym_N}\sgn(pp')
s_{-i_{p'(1)},i_{p(1)}}(u-N+1)\cdots
s_{-i_{p'(n)},i_{p(n)}}(u-N+n)&\\
\cdot s^t_{-i_{p'(n+1)},i_{p(n+1)}}(-u+N-n-1)\cdots
s^t_{-i_{p'(N)},i_{p(N)}}(-u)&,
\endaligned\tag2
$$
where $s^t_{ij}(u)$ are the matrix elements of the matrix $S^t(u)$.
\endproclaim

\bigskip
In particular, if $N=2$ and $(i_1,i_2)$ is a permutation of the indices
$(-1,1)$ then replacing $s_{ij}^t(u)$ with $\theta_{ij}s_{-j,-i}(u)$ we
get
$$
\align
\sdet S(u)&={2u+1\over 2u\pm 1}(s_{-i_1,i_2}(-u)s_{-i_1,i_2}(u-1)\mp
s_{-i_1,i_1}(-u)s_{-i_2,i_2}(u-1))\\
&={2u+1\over 2u\pm 1}(s_{-i_1,i_2}(u-1)s_{-i_1,i_2}(-u)\mp
s_{-i_1,i_1}(u-1)s_{-i_2,i_2}(-u)),
\endalign
$$
while rewriting (2.6.1) in terms of the matrix
elements directly, we can obtain the expressions
$$
\align
&\sdet S(u)\\
&=s_{-i_2,i_1}(u)s_{-i_1,i_2}(u-1)-s_{-i_1,i_1}(u)s_{-i_2,i_2}(u-1)\\
&\qquad\qquad\qquad\qquad\qquad\qquad\qquad\qquad
+{1\over 2u-1}(s_{-i_2,i_1}(u)\mp s_{-i_1,i_2}(u))s_{-i_1,i_2}(u-1)\\
&=s_{-i_1,i_2}(u-1)s_{-i_2,i_1}(u)-s_{-i_1,i_1}(u-1)s_{-i_2,i_2}(u)\\
&\qquad\qquad\qquad\qquad\qquad\qquad\qquad\qquad
+{1\over 2u-1}s_{-i_1,i_2}(u-1)(s_{-i_2,i_1}(u)\mp s_{-i_1,i_2}(u)).
\endalign
$$
They coincide with the previous ones due to the symmetry relation (3.3),
which can be written in this case as
$$
\frac{2u}{2u-1}s_{-i_2,i_1}(u)\mp\frac1{2u-1}
s_{-i_1,i_2}(u)=\frac{2u+1}{2u\pm1}s_{-i_1,i_2}(-u)
$$
and
$$
s_{-i_a,i_a}(u)=\pm\frac{2u+1}{2u\pm1}s_{-i_a,i_a}(-u), \quad a=1,2.
$$
If $N=3$ and $(i_1,i_2,i_3)$ is a permutation of the indices
$(-1,0,1)$ then
$$
\align
\sdet &S(u)\\
&=s_{-i_2,i_1}(-u)s_{-i_2,i_1}(u-1)s_{-i_3,i_3}(u-2)
-s_{-i_2,i_1}(-u)s_{-i_3,i_1}(u-1)s_{-i_2,i_3}(u-2)\\
&-s_{-i_2,i_2}(-u)s_{-i_1,i_1}(u-1)s_{-i_3,i_3}(u-2)
-s_{-i_1,i_2}(-u)s_{-i_3,i_2}(u-1)s_{-i_1,i_3}(u-2)\\
&+s_{-i_2,i_3}(-u)s_{-i_1,i_1}(u-1)s_{-i_2,i_3}(u-2)
+s_{-i_1,i_3}(-u)s_{-i_2,i_2}(u-1)s_{-i_1,i_3}(u-2)
\endalign
$$
and also
$$
\align
&=s_{-i_2,i_1}(u-2)s_{-i_2,i_1}(-u+1)s_{-i_3,i_3}(-u)
-s_{-i_2,i_1}(u-2)s_{-i_3,i_1}(-u+1)s_{-i_2,i_3}(-u)\\
&-s_{-i_2,i_2}(u-2)s_{-i_1,i_1}(-u+1)s_{-i_3,i_3}(-u)
-s_{-i_1,i_2}(u-2)s_{-i_3,i_2}(-u+1)s_{-i_1,i_3}(-u)\\
&+s_{-i_2,i_3}(u-2)s_{-i_1,i_1}(-u+1)s_{-i_2,i_3}(-u)
+s_{-i_1,i_3}(u-2)s_{-i_2,i_2}(-u+1)s_{-i_1,i_3}(-u).
\endalign
$$
The direct application of (2.6.1) gives in this case a formula
with 48 summands.

\bigskip
\Proof We shall prove the theorem in several steps.

{\sl Step} 1. Applying relations (3.4.4)--(3.4.8) and
Proposition 3.5 we obtain
the following recurrent formula:
$$
\aligned
m(m-1)\frac{2u\pm1}{2u+1}\ts
\si_{i_1,\dots,i_{m-1},i_m}^{-i_1,\dots,-i_{m-1},-j_m}&(u)=\\
\sum_{k,l=1;\ k\ne l}^{m-1}s_{-i_ki_l}^t(-u)\ts&
\si_{i_1,\dots,\widehat{i_k},\dots,\widehat{i_l},\dots,i_{m-1},i_k}
^{-i_1,\dots,-\widehat{i_k},\dots,-\widehat{i_l},\dots,-i_{m-1},-j_m}(u-1)\t
s
s_{-i_li_m}(u-m+1)\\
-\sum_{k=1}^{m-2}s_{-i_ki_{m-1}}^t(-u)\ts&
\si_{i_1,\dots,\widehat{i_k},\dots,i_{m-2},i_k}
^{-i_1,\dots,-\widehat{i_k},\dots,-i_{m-2},-i_{m-1}}(u-1)\ts
s_{-j_mi_m}(u-m+1)\\
-\sum_{l=1}^{m-1}s_{-j_mi_l}^t(-u)\ts&
\si_{i_1,\dots,\widehat{i_l},\dots,i_{m-1}}
^{-i_1,\dots,-\widehat{i_l},\dots,-i_{m-1}}(u-1)\ts
s_{-i_li_m}(u-m+1)\\
+s_{-i_{m-1}i_{m-1}}^t(-u)\ts&
\si_{i_1,\dots,i_{m-2}}
^{-i_1,\dots,-i_{m-2}}(u-1)\ts
s_{-j_mi_m}(u-m+1).
\endaligned\tag3
$$
Now let $(i_1,\dots,i_N)$ be a fixed permutation of the indices
$(-n,\dots,n)$.
Denote by $q_0$ the element of $\Sym_N$ such that
$i_{q_0(k)}=-i_k$ for $k=1,\dots,N$. Clearly, $\sgn(q_0)=(-1)^n$. Hence,
formula (3.4.3) gives
$$
\sdet S(u)=(-1)^n\ts N!\ts\si_{i_1,\dots,i_N}
^{-i_1,\dots,-i_N}(u).
$$
Let us verify that starting with the
element $\dsize\si_{i_1,\dots,i_N}^{-i_1,\dots,-i_N}(u)$,
and applying formula (3) repeatedly,
we get formula (1). First we note that
$$
\frac{2u\pm1}{2u+1}\cdot\frac{2(u-1)\pm1}{2(u-1)+1}\cdots
\frac{2(u-n+1)\pm1}{2(u-n+1)+1}=\gamma_N^{-1}(u).
$$
So, after the $n$-th step the expression $\dsize\gamma_N^{-1}(u)\ts N!\ts
\si_{i_1,\dots,i_N}^{-i_1,\dots,-i_N}(u)$ will be represented as a
linear combination of the products of the form
$$
\aligned
s^t_{-i_{p(1)},i_{q(1)}}(-u)\cdots
&s^t_{-i_{p(n)},i_{q(n)}}(-u+n-1)\\
\cdot &s_{-i_{p(n+1)},i_{q(n+1)}}(u-n)\cdots
s_{-i_{p(N)},i_{q(N)}}(u-N+1),\qquad p,q\in\Sym_N
\endaligned\tag4
$$
with coefficients equal $1$ or $-1$. It is clear that $q(N)=N$.
Furthermore, we see from formula (3), applied to the element
$\dsize\si_{i_1,\dots,i_N}^{-i_1,\dots,-i_N}(u)$ that for each
product (4) the following conditions hold:
$$
\alignat2
&\text{if}\quad p(1),\ p(N)<N\qquad\qquad&&\text{then}\quad
q(1)=p(N),\ q(N-1)=p(1),\\
&\text{if}\quad p(1)<N-1,\ p(N)=N\qquad\qquad&&\text{then}\quad q(1)=N-1,
\ q(N-1)=p(1),\\
&\text{if}\quad p(1)=N,\ p(N)<N-1\qquad\qquad&&\text{then}\quad
q(1)=p(N),\ q(N-1)=N-1,\\
&\text{if}\quad p(1)=N-1,\ p(N)=N\qquad\qquad&&\text{then}\quad q(1)=N-1,
\ q(N-1)=N-2,\\
&\text{if}\quad p(1)=N,\ p(N)=N-1\qquad\qquad&&\text{then}\quad
q(1)=N-1,\ q(N-1)=N-2.
\endalignat
$$
Applying now formula (3) to the element
$$
\si_{i_1,\dots,\widehat i_{p(1)},\dots,\widehat i_{p(N)},\dots,i_{q(N-1)}}
^{-i_1,\dots,-\widehat i_{p(1)},\dots,-\widehat i_{p(N)},\dots,-i_N}(u-1)
$$
we find that essentially the same conditions hold for the numbers
$p(2),p(N-1),q(2),q(N-2)$, considered as elements of the set
$\{1,\dots,\widehat{p(1)},\dots,\widehat{p(N)},\dots,N\}$, etc. Comparing
this
with the transformation described
in Subsection 3.1, we conclude that $q$ coincides with
the permutation $p'$ given by (3.1.2).

{\sl Step} 2. Now we prove that the coefficient of the product (4) equals
$\sgn(pp')$. Using the result of Step 1, we can write
$$
\aligned
N!&\ts\frac{2u\pm1}{2u+1}\ts
\si_{i_1,\dots,i_N}^{-i_1,\dots,-i_N}(u)=\\
&\sum_{p\in\Sym_N}\varepsilon(p)s_{-i_{p(1)},i_{p'(1)}}^t(-u)\ts
\si_{i_1,\dots,\widehat i_{p(1)},\dots,\widehat i_{p(N)},\dots,i_{p'(N-1)}}
^{-i_1,\dots,-\widehat i_{p(1)},\dots,-\widehat
i_{p(N)},\dots,-i_N}(u-1)\ts
s_{-i_{p(N)},i_N}(u-N+1),
\endaligned\tag5
$$
where
$$
\varepsilon(p)=\cases \ \ 1,\quad&\text{if\ \ either}\quad p(1),\ p(N)<N,
\quad\text{or}\quad p(1)=N-1,\ p(N)=N;\\
-1,\quad&\text{if\ \ either}\quad p(1)<N-1,\ p(N)=N,
\quad\text{or}\quad p(1)=N,\ p(N)<N.\endcases
$$
\bigskip
\noindent
For any sequence of distinct positive integers $a=(a_1,\dots,a_m)$
denote by $\inv(a)$ the number of inversions in $a$.
For a permutation $p\in\Sym_N$ we let $\tilde p$ and $\tilde p'$ denote
the sequences $(p(2),\dots,p(N-1))$ and $(p'(2),\dots,p'(N-2))$
respectively. It suffices to check the following identity:
$$
(-1)^{\inv(p)}\cdot(-1)^{\inv(p')}=\varepsilon(p)\ts
(-1)^{\inv(\tilde p)}\cdot(-1)^{\inv(\tilde p')}.\tag6
$$
Then the required statement will follow from (5) with the use of
the obvious induction argument.

Suppose that $p(1)=k$, $p(N)=l$. If $k,l<N$ then clearly
$\varepsilon(p)=1$ while
$$
\align
&\inv(p)=(k-1)+(N-l)+\inv(\tilde p),\\
&\inv(p')=(l-1)+(N-k-2)+\inv(\tilde p')\quad\text{if}\quad k<l;
\endalign
$$
and
$$
\align
&\inv(p)=(k-1)+(N-l-1)+\inv(\tilde p),\\
&\inv(p')=(l-1)+(N-k-1)+\inv(\tilde p')\quad\text{if}\quad k>l;
\endalign
$$
so that in both cases (6) holds. Similarly, if $k<N-1$, $l=N$
then
$$
\align
&\varepsilon(p)=-1,\\
&\inv(p)=(k-1)+\inv(\tilde p),\\
&\inv(p')=(N-2)+(N-k-2)+\inv(\tilde p');
\endalign
$$
if $k=N$, $l<N-1$ then
$$
\align
&\varepsilon(p)=-1,\\
&\inv(p)=(N-1)+(N-l-1)+\inv(\tilde p),\\
&\inv(p')=(l-1)+\inv(\tilde p');
\endalign
$$
if $k=N-1$, $l=N$ then
$$
\align
&\varepsilon(p)=1,\\
&\inv(p)=(N-2)+\inv(\tilde p),\\
&\inv(p')=(N-2)+\inv(\tilde p');
\endalign
$$
if $k=N$, $l=N-1$ then
$$
\align
&\varepsilon(p)=-1,\\
&\inv(p)=(N-1)+\inv(\tilde p),\\
&\inv(p')=(N-2)+\inv(\tilde p').
\endalign
$$
This proves (6) and formula (1).

{\sl Step} 3. To prove formula (2) we use relation (2.7.1).
We obtain from (1) that
$$
\align
\sdet S(u)=\qquad\qquad\qquad\qquad\qquad&\\
(-1)^n\gamma_N(u)\sum_{p\in\Sym_N}\sgn(pp')
&s^t_{-i_{p(1)},i_{p'(1)}}(u-N+1)\cdots
s^t_{-i_{p(n)},i_{p'(n)}}(u-N+n)\\
\cdot &s_{-i_{p(n+1)},i_{p'(n+1)}}(-u+N-n-1)\cdots
s_{-i_{p(N)},i_{p'(N)}}(-u).
\endalign
$$
Note that
$$
s^t_{-i_{p(k)},i_{p'(k)}}(u-N+k)=
\theta_{-i_{p(k)},i_{p'(k)}} s_{-i_{p'(k)},i_{p(k)}}(u-N+k),
\qquad k=1,\dots,n
$$
and
$$
s_{-i_{p(k)},i_{p'(k)}}(-u+N-k)
=\theta_{-i_{p(k)},i_{p'(k)}}
s^t_{-i_{p'(k)},i_{p(k)}}(-u+N-k),\qquad k=n+1,\dots,N.
$$
Now (2) follows from the equality
$$
\prod_{k=1}^N\theta_{-i_{p(k)},i_{p'(k)}}=1.
$$
The theorem is proved.

\newpage
\heading
{\bf 4. Generators of the center of the algebra $\U(\g(n))$}
\endheading
\

Here we apply Theorem 3.6 to the construction of a system of free
generators of the center $\Z(\g(n))$ of the algebra $\U(\g(n))$.

\bigskip
\noindent {\bf 4.1.}
Let us set for $i=1,\dots,n$
$$
\rho_{-i}=-\rho_i=\cases i-1,&\text{for$\quad\g(n)=\oa(2n),$}\\
i-\tfrac12,&\text{for$\quad\g(n)=\oa(2n+1),$}\\
i,&\text{for$\quad\g(n)=\spa(2n)$}.\endcases
$$
We also set $\rho_0:=1/2$ in the case of $\g(n)=\oa(2n+1)$.
Let us apply the homomorphism (2.5.2) to the series
$\sdet S(u+N/2-1/2)$ and set
$$
C(u):=\gamma_N^{-1}(u+N/2-1/2)\prod_{i=-n}^n(u+\rho_i)\thinspace
\xi(\sdet S(u+N/2-1/2)) \tag1
$$
(index $0$ is skipped in the product if $N=2n$).

\bigskip
\proclaim
{\bf Proposition} $C(u)$ is a monic polynomial in $u$ of degree $N$ with
coefficients from $\Z(\g(n))$ and it can be written as
$$
C(u)=(-1)^n\sum_{p\in\Sym_N}\sgn(pp')
(u+\rho_{-n}+F)_{-i_{p(1)},i_{p'(1)}}\cdots
(u+\rho_{n}+F)_{-i_{p(N)},i_{p'(N)}}, \tag2
$$
where $(i_1,\dots,i_N)$ is an arbitrary
permutation of the indices \newline $(-n,-n+1,\dots,n)$
and $p'$ is the image of $p$
under the projection {\rm (3.1.2)}.
\endproclaim

\Proof Theorem 2.8 immediately implies that all the coefficients of the
series
$C(u)$ belong to $\Z(\g(n))$.
It follows from relation (2.5.1) that
$$
\xi(s^t_{ij}(-v))=\delta_{ij}+(F^t)_{ij}(-v\pm\frac12)^{-1}
=\delta_{ij}+F_{ij}(v\mp\frac12)^{-1}.
$$
Hence,
$$
(v\mp\frac12)\xi(s^t_{ij}(-v))=F_{ij}(v\mp1),
$$
and using a simple calculation we derive
formula (2) from (3.6.1). This proves that
$C(u)$ is a polynomial in $u$ of
degree $\leq N$.

Suppose now that $(i_1,\dots,i_N)$ is the identical
permutation of the indices $(-n,-n+1,\dots,n)$ and prove that
there exists the only element $p\in\Sym_N$ such that
$$
-i_{p(k)}=i_{p'(k)}\quad\text{for every}\quad k=1,\dots,N. \tag 3
$$
Indeed, $i_{p'(N)}=i_N=n$, hence $i_{p(N)}=-n$ and so, $p(N)=1$. Further,
we obtain from the definition of the map $p\ra p'$, that $p'(1)=1$ and
(3) implies that $p(1)=N$. In this case $p'(N-1)=N-1$. Using an easy
induction, we conclude that
$$
p=(N,N-1,\dots,1)\quad\text{and}\quad p'=(1,2,\dots,N).\tag4
$$
Clearly, $\sgn(pp')=(-1)^n$ and so, we see from (1) and (2) that
the leading term of $C(u)$ is $u^N$, which proves the proposition.

\bigskip
\noindent {\bf 4.2.}
In the case of $\g(n)=\oa(2n)$ introduce the element
$z'_n\in\U(\oa(2n))$ as follows:
$$
z'_n=\sum_{p\in\Sym_{2n}}\sgn(p)F_{-i_{p(1)},i_{p(2)}}\cdots
F_{-i_{p(2n-1)},i_{p(2n)}},
$$
where $(i_1,\dots,i_{2n})=(-n,\dots,n)$.
Then $z'_n$ belongs to the center of the algebra $\U(\oa(2n))$ (see, for
instance, [Z]).

\proclaim
{\bf Theorem} The polynomial $C(u)$ has the form
$$
C(u)=u^{2n}+z_2u^{2n-2}+\cdots+z_{2n},\quad z_{2i}\in\Z(\g(n))\tag1
$$
for $N=2n$, and
$$
C(u)=(u+\frac12)(u^{2n}+z_2u^{2n-2}+\cdots+z_{2n}),\quad z_{2i}\in\Z(\g(n))
\tag2
$$
for $N=2n+1$.
Moreover, the elements $z_2,\dots,z_{2n}$ are free generators of
$\Z(\g(n))$ for $\g(n)=\oa(2n+1),\ \spa(2n)$ and
the elements $z_2,\dots,z_{2n-2},z'_n$ are those for $\g(n)=\oa(2n)$.
\endproclaim

\Proof Formula (2.7.1) implies that if $N=2n$ then
$$
C(u)=C(-u),\tag3
$$
and if $N=2n+1$ then
$$
(-u+\frac12)\ts C(u)=(u+\frac12)\ts C(-u),\tag4
$$
which proves formulae (1) and (2).

To prove the second part of the theorem we find the images of
the elements $z_{2i}$ under
the Harish-Chandra homomorphism. Denote by $\n_+$, $\n_-$ and $\h$
the subalgebras in $\g(n)$ spanned by the elements
$F_{ij}$ with $i<j$, $i>j$ and $i=j$ respectively.
We identify the algebra $\U(\h)$ with the algebra of polynomial
functions on $\h^*$ and denote by $\lambda_i$ the function which
corresponds to $F_{ii}$. For $i=1,\dots,n$ set
$l_{i}=-l_{-i}:=\lambda_{i}+\rho_i$ and in the case of $\g(n)=\oa(2n+1)$
set $l_0:=1/2$.
Consider formula (4.1.2)
with $(i_1,\dots,i_N)$ being the identical
permutation of the indices $(-n,-n+1,\dots,n)$. Since $i_{p'(N)}=i_N=n$
we see that
$$
(u+\rho_n+F)_{-i_{p(N)},i_{p'(N)}}=F_{-i_{p(N)},i_{p'(N)}}\in\n_+
$$
unless $i_{p(N)}=-n$,
that is, $p(N)=1$. In the latter case $p'(1)=1$ and so,
$$
(u+\rho_{-n}+F)_{-i_{p(1)},i_{p'(1)}}=F_{-i_{p(1)},i_{p'(1)}}\in\n_-
$$
unless $i_{p(1)}=n$, that is, $p(1)=N$. This implies that $p'(N-1)=N-1$
and we can repeat the same arguments for the elements
$$
(u+\rho_{n-1}+F)_{-i_{p(N-1)},i_{p'(N-1)}}\quad\text{and}
\quad (u+\rho_{-n+1}+F)_{-i_{p(2)},i_{p'(2)}},
$$
etc.\ and prove that the image $C_{\lambda}(u)$ of the polynomial $C(u)$
under
the Harish-Chandra homomorphism is determined by the only summand in
(4.1.2) with
$p$ and $p'$ given by (4.1.4). By (2.5.1), $F_{-i,-i}=-F_{ii}$, and so,
$$
C_{\lambda}(u)=(u^2-l_{1}^2)\cdots (u^2-l_{n}^2)\tag 5
$$
for $N=2n$, and
$$
C_{\lambda}(u)=(u+\frac12)(u^2-l_{1}^2)\cdots (u^2-l_{n}^2)\tag 6
$$
for $N=2n+1$. Hence, the images of the elements $z_{2i}$, $i=1,\dots,n$
are the elementary symmetric polynomials in the
variables $-l_{1}^2,\dots,-l_{n}^2$.
It is well known that the image of the element $z'_n$ under the
Harish-Chandra homomorphism is, up to
a constant, the polynomial $l_{1}\cdots l_{n}$ [Z].
So, in each of the three cases we obtain that the images of the
elements $z_2,\dots,z_{2n}$ or $z_2,\dots,z_{2n-2},z'_n$ are
algebraically independent polynomials, and examining their degrees we
conclude that these elements are free generators of the algebra
$\Z(\g(n))$.
The theorem is proved.

Note that, as it follows from the proof, in the case of
$\g(n)=\oa(2n)$
the central elements $(z'_n)^2$ and $z_{2n}$ are proportional.

\newpage
\heading
{\bf 5. Characteristic identities for the matrix $F$}
\endheading
\

In this section we prove an analogue of the Cayley--Hamilton theorem
for the matrix $F$ (cf. Theorem 1.6) and show that it implies the
polynomial identities obtained by
Bracken and Green [BG], [Gr].

\bigskip
\noindent {\bf 5.1.} Consider the polynomial $C(u)$ introduced in
Section 4.
\bigskip
\proclaim
{\bf Theorem} The matrix $F$ satisfies the polynomial
identity:
$$
C(-F-\rho_n)=0.\tag1
$$
\endproclaim

\Proof Let us define the matrix $\widehat S(u)=
(\widehat s_{ij}(u))$ by the formula
$$
\align
A_NS_1(u)R'_{12}\cdots R'_{1N}S_2(u-1)&
R'_{23}\cdots R'_{2N}S_3(u-2)\\
\cdots &S_{N-1}(u-N+2)R'_{N-1,N}
=A_N\widehat S_N(u).
\endalign
$$
Then, by (3.4.2)
$$
\widehat s_{i_Nj}(u)=
N!\ts\tsi_{i_1,\dots,i_{N-1},j}^{i_1,\dots,i_N}(u),
$$
where $(i_1,\dots,i_N)$ is a permutation
of the indices $(-n,-n+1,\dots,n)$ and $j\in\{-n,\dots,n\}$.
Relations (3.4.6) and (3.4.8) imply that the elements
$\widehat s_{ij}(u)$ are
well-defined. Furthermore, by Theorem 2.6
$$
A_N\widehat S_N(u)S_N(u-N+1)=A_N\ts\sdet S(u)
$$
and hence,
$$
\sdet S(u)=\widehat S(u)\ts S(u-N+1).\tag 2
$$
Let us replace $u$ with $u+N/2-1/2$ in this formula and apply the
homomorphism $\xi$ (see Theorem 2.5) to both of its sides.
Since
$$
\xi(S(u-N/2+1/2))=1+F(u+\rho_n)^{-1}
$$
we get the equality
$$
\xi(\sdet S(u+N/2-1/2))=
\xi(\widehat S(u+N/2-1/2))(1+F(u+\rho_n)^{-1}).
$$
Multiplying now both sides by the product
$\gamma_N(u+N/2-1/2)^{-1}(u+\rho_{-n})\cdots
(u+\rho_{n})$
and using Proposition 3.5 and formula (3.6.3) we obtain that
$$
C(u)=\widehat C(u)\thinspace(u+\rho_n+F),
$$
where $\widehat C(u)$ is a polynomial in $u$ with coefficients from
$\U(\g(n))\otimes\End\E$. This proves
the theorem.

\bigskip
\noindent {\bf 5.2.}
Let $L(\lambda)$, $\lambda=(\lambda_{-n},\dots,\lambda_{-1})\in\h^*$ denote
the
highest weight representation of the Lie algebra $\g(n)$, that is,
$L(\lambda)$ contains a nonzero vector $v$ such that $L(\lambda)$ is
generated \newline
by $v$,
$$
\n_+\ts v=0, \qquad\text{and}\qquad F_{-i,-i}\ts v=\lambda_{-i}\ts v\quad
\text{for}\quad i=1,\dots,n.
$$
We have the following analogue of Corollary 1.7 (cf. [BG], [Gr]).

\bigskip
\proclaim
{\bf Corollary} The image of the matrix $F$ in
the representation $L(\lambda)$ satisfies the identity
$$
\prod_{i=-n}^n(F+\rho_n-l_i)=0.\tag1
$$
\endproclaim

\Proof It is clear that the image of
the polynomial $C(u)$ in the representation
$L(\lambda)$ coincides with the polynomial $C_{\lambda}(u)$ which
was defined in
the proof of Theorem 4.2 (see formulae (4.2.5) and (4.2.6)). Performing
the obvious calculation we find that the image of (5.1.1) in $L(\lambda)$
has the required form. The proof is completed.

\bigskip
\noindent
{\bf Note.} I would like to thank M. Nazarov
who communicated me that he had
managed to derive the identity (5.2.1) by using
the quantum Liouville formula (see [N], [MNO]) and
the connection between the Sklyanin determinant and the quantum
determinant (Theorem 2.7).

\newpage
\heading
{\bf References}
\endheading
\bigskip

\itemitem{[BG]}
{A. J. Bracken and H. S. Green},
{\sl Vector operators and a polynomial identity for SO(n)},
{J. Math. Phys.}
{\bf 12}
(1971),
2099--2106.

\itemitem{[BCC]}
{D. M. O'Brien, A. Cant and A. L. Carey},
{\sl On characteristic identities for Lie algebras},
{Ann. Inst. Henri Poincar\'e}
{\bf A26}
(1977),
405--429.

\itemitem{[Ca1]}
{A. Capelli},
{\sl \"Uber die Zur\"uckf\"uhrung der Cayley'schen Operation
$\Omega$ auf ge\-w\"ohn\-lich\-en Polar-Operationen}, {Math. Ann.}
{\bf 29}
(1887),
331--338.

\itemitem{[Ca2]}
{A. Capelli},
{\sl Sur les op\'erations dans la th\'eorie des formes
alg\'ebriques},
{Math. Ann.}
{\bf 37}
(1890),
1--37.

\itemitem{[C]}
{I. V. Cherednik},
{\sl A new interpretation of Gelfand--Tzetlin bases}, {Duke Math. J.}
{\bf 54}
(1987),
563--577.

\itemitem{[D]}
{V. G. Drinfeld},
{\sl Hopf algebras and the quantum Yang--Baxter equation}, {Soviet Math.
Dokl.}
{\bf 32}
(1985),
254--258.

\itemitem{[G]}
{M. D. Gould},
{\sl Characteristic identities for semi-simple Lie algebras},
{J. Austral. Math. Soc.}
{\bf B26}
(1985),
257--283.

\itemitem{[Gr]}
{H. S. Green},
{\sl Characteristic identities for generators of GL(n), O(n) and Sp(n)},
{J. Math. Phys.}
{\bf 12}
(1971),
2106--2113.

\itemitem{[H]}
{R. Howe},
{\sl Remarks on classical invariant theory}, {Trans. AMS}
{\bf 313}
(1989),
539--570.

\itemitem{[HU]}
{R. Howe and T. Umeda},
{\sl The Capelli identity, the double commutant theorem,
and multiplicity-free actions},
{Math. Ann.}
{\bf 290}
(1991),
569--619.

\itemitem{[KS]}
{P. P. Kulish and E. K. Sklyanin},
{\sl Quantum spectral transform method: recent developments},
in \lq Integrable Quantum Field Theories', {Lecture Notes in Phys.}
{\bf 151}
Springer,
Berlin-Heidelberg,
1982,
pp. 61--119.

\itemitem{[M]}
{A. I. Molev},
{\sl Gelfand--Tsetlin bases for representations of Yangians},
{Lett. Math. Phys.},
{\bf 30}
(1994),
53--60.

\itemitem{[MNO]}
{A. I. Molev, M. L. Nazarov and G. I. Olshanski\u\i},
{\sl Yangians and classical Lie algebras},
{Preprint CMA-MR53-93, Austral. Nat. Univ.}, November 1993.
(hep-th$\backslash9409025$)

\itemitem{[N]}
{M. L. Nazarov},
{\sl Quantum Berezinian and the classical Capelli identity},
{Lett. Math. Phys.}
{\bf 21}
(1991),
123--131.

\itemitem{[NT]}
{M. Nazarov and V. Tarasov},
{\sl Yangians and Gelfand--Zetlin bases},
Preprint RIMS, Kyoto Univ.,
February 1993 (to appear in Publications of RIMS, {\bf 30} (1994), no 3).

\itemitem{[O1]}
{G. I. Olshanski\u\i},
{\sl Representations of infinite-dimensional classical
groups, limits of enveloping algebras, and Yangians},
in \lq Topics in Representation Theory (A. A. Kirillov, Ed.), {Advances in
Soviet Math.} {\bf 2},
AMS,
Providence RI,
1991,
pp. 1--66.

\itemitem{[O2]}
{G. I. Olshanski\u\i},
{\sl Twisted Yangians and infinite-dimensional classical Lie algebras},
in \lq Quantum Groups (P. P. Kulish, Ed.)', {Lecture Notes in Math.}
{\bf 1510},
Springer,
Berlin-Heidelberg,
1992,
pp. 103--120.

\itemitem{[PP]}
{A. M. Perelomov and V. S. Popov},
{\sl Casimir operators for semisimple Lie algebras}, {Isv. AN SSSR}
{\bf 32}
(1968),
1368--1390.

\itemitem{[RTF]}
{N. Yu. Reshetikhin, L. A. Takhtajan and L. D. Faddeev},
{\sl Quantization of Lie Groups and Lie algebras}, {Leningrad Math. J.}
{\bf 1}
(1990),
193--225.

\itemitem{[S]}{R. P. Stanley},
{\sl Enumerative combinatorics, I},
(Wadsworth and Brooks/Cole, Monterey, California
1986).

\itemitem{[TF]}
{L. A. Takhtajan and L. D. Faddeev},
{\sl Quantum inverse scattering method and the Heisenberg
$XYZ$-model},
{Russian Math. Surv.}
{\bf 34}
(1979),
no. 5,
11--68.

\itemitem{[W]}
{H. Weyl},
{\sl Classical Groups, their Invariants and Representations},
{Princeton Univ. Press},
Princeton NJ,
1946.

\itemitem{[Z]}
{D. P. Zhelobenko},
{\sl Compact Lie groups and their representations},
{Transl. of Math. Monographs}
{\bf 40} AMS,
Providence RI
1973.

\enddocument
\end